\documentclass{aa}
\usepackage{graphicx,times}

\begin{document}
\DeclareGraphicsExtensions{.pdf,.png,.gif,.jpg}

\title{Temperature and entropy profiles of nearby cooling flow clusters observed with XMM-Newton}

\author{ R. Piffaretti \inst{1,2}
         \and
         Ph. Jetzer \inst{1}
         \and
         J.S. Kaastra \inst{3}
         \and
         T. Tamura \inst{3,4}                 
       }
  
\offprints{R. Piffaretti}
\mail{piff@physik.unizh.ch}

\institute{Institute of Theoretical Physics, University of Z\"{u}rich, Winterthurerstrasse, 190, CH-8057 Z\"{u}rich, Switzerland
              \and
              Laboratory for Astrophysics,
              Paul Scherrer Institute,  CH-5232 Villigen PSI, Switzerland
             \and
              SRON National Institute for Space Research,
              Sorbonnelaan 2, 3584 CA Utrecht, The Nether\-lands
              \and
              Institute of Space and Astronautical Science, 3-1-1 Yoshinodai, Sagamihara, Kanagawa 229-8510, Japan 
              }
\date{Received  / Accepted  }

\abstract{We investigate temperature and entropy profiles of 13 
nearby cooling flow clusters observed with the EPIC cameras of
XMM-Newton. When normalized
and scaled by the virial radius the temperature profiles turn out to
be remarkably similar. At large radii the temperature profiles show a
clear decline starting from a break radius at $\sim 0.1 \, r_{\mathrm{vir}}$. The
temperature decreases by $\sim 30 \%$ between $0.1  \, r_{\mathrm{vir}}$ and
$0.5 \, r_{\mathrm{vir}}$. As expected for systems where non-gravitational processes
are of great importance, the scale length characterizing the central temperature
drop is not found to be proportional to the virial radius of the system. The
entropy of the plasma increases monotonically moving outwards almost proportional to
the radius and the central entropy level is tightly correlated with
the core radius of the X-ray emission. The dispersion in the entropy profiles is smaller if the
empirical relation $S \propto T^{0.65}$ is used
instead of the standard self-similar relation $S \propto T$ and, as expected for cooling flow clusters, no entropy cores
are observed.

}
\maketitle

\section{Introduction}

Temperature profiles of the X-ray-emitting intracluster
medium (ICM) have several important implications. Derived from X-ray
data together with the gas density distribution, they are critical in
deriving the total gravitational mass of galaxy clusters, which
combined with the estimated gas mass and primordial nucleosynthesis
calculations, provides constraints on the cosmological density
parameter (e.g., \cite{white93}). Temperature profiles provide useful knowledge to complementary
observations such as the Sunyaev-Zel'dovich effect and affect the determination of
cosmological parameters like the Hubble constant (e.g.,
\cite{rephaeli1995}). They also provide valuable information on the
thermodynamic history of galaxy clusters, since they are fundamental in determining the ICM
entropy distribution, and the latter has been
shown to be a very powerful tool to study non-gravitational processes such as radiative
cooling, preheating and feedback from supernovae and active 
 galactic nuclei (e.g., \cite{voitrev}).\\*
Various observational studies have found different and conflicting results regarding
temperature gradients in the outer regions of galaxy clusters. \cite{markevitch} analyzed projected radial
temperature profiles for 30 clusters observed with ASCA, finding that
nearly all the clusters in their sample, including cooling flow
clusters, show a significant decline at large radii. \cite{irwin}
pointed out that the claim by \cite{markevitch} of a
declining temperature profile is in contrast with the findings of
other authors even for the same object and strengthened the claim of
isothermality by investigating ROSAT PSPC color profiles. \cite{white} studied a sample of 106 clusters observed with ASCA and found that 90
percent of the temperature profiles are consistent with isothermality.
\cite{irwin00} derived radial temperature profiles for 11
clusters observed with {\it BeppoSAX} and found generally flat or slightly
increasing temperature profiles, therefore ruling out a declining temperature
profile. On the other hand \cite{finoguenov01} found declining
profiles similar to those of \cite{markevitch} from the
analysis of ASCA/SIS data of 18 relaxed cool clusters. \cite{dm02} analyzed the projected temperature	
profiles for 11 cooling flow and 10 non-cooling flow clusters observed
with {\it BeppoSAX}. They find a rapid decline of the profiles beyond $\sim
 0.2 \,  r_{\mathrm{180}}$ and that non-cooling
flow clusters appear to be well described by an isothermal core
extending out to $\sim 0.2 \, r_{\mathrm{180}}$, while the well known
signature of cooling flow clusters is the temperature decline towards
the center. Despite the agreement between \cite{markevitch}, \cite{finoguenov01} and \cite{dm02}, the behavior of cluster temperature profiles
beyond $\sim 0.2 r_{180}$ remains controversial, mainly because even
with large samples the number of clusters where the temperature is
derived beyond $0.5 r_{180}$ is very small.\\*
In the early phases of the
XMM-Newton mission, \cite{arnaud01} found that the temperature profile of A~1795 is
essentially isothermal up to $\sim 0.4 \, r_{\mathrm{200}}$. \cite{pratt01} find a flat temperature profile up to $\sim
0.5 \, r_{\mathrm{200}}$ for A~2163 and an indication of a mild decline at larger
radii. XMM-Newton observation of the distant cluster RX J1120.1+4318
indicate a flat profile up to $\sim 0.5 \, r_{\mathrm{vir}}$ (\cite{arnaud02}), and no sharp temperature drop is found in A~1835 up to $\sim 0.7
\, r_{\mathrm{vir}}$ (\cite{majerowicz02}). \cite{pratt02} find a declining temperature profile in
A~1413. The gradient is modest compared to the results of
\cite{markevitch} and \cite{dm02}: the temperature decreases by $\sim 20 \%$
between $0.1 \, r_{\mathrm{200}}$ and $0.5 \, r_{\mathrm{200}}$. The
temperature profile of A~1983 derived within $\sim 0.4 \,
r_{\mathrm{200}}$ by \cite{pratt03} does not show a decline and for A~1650
\cite{takahashi} find a temperature decline of $\sim 20 \%$ between
$0.1 \, r_{\mathrm{200}}$ and $0.4 \, r_{\mathrm{200}}$. Given these controversial results on
temperature gradients at large radii, any further insight is welcome.\\*
In this paper we thus investigate radial temperature profiles of nearby
cooling flow clusters using the results from the spatially resolved
spectra taken with the EPIC cameras of XMM-Newton presented in
\cite{kaastra03}. We present an accurate description of temperature
and entropy profiles at all the observed radii and investigate their
scaling properties. We additionally provide information on the
``uncorrected'' luminosity-temperature relation for cooling flow
clusters.\\*
The layout of this paper is as follows. In Sect.~\ref{sect:sample} we
briefly summarize the procedure and results which are essential for
our analysis (see \cite{kaastra03} for the description of the sample and the extensive presentation of the
data analysis) and discuss the selection of temperature bins. 
In Sect.~\ref{sect:masses} we compute, for each cluster, gas and total
gravitational masses, and the
virial radius needed for the scaling of temperature profiles presented in
Sect.~\ref{sect:Tprofiles} along with the comparison with previous
work and numerical simulations. In Sect.~\ref{sect:entropy} we focus
on the scaling of entropy profiles and present the luminosity
temperature relation derived from our sample. 
In Sect.~\ref{sect:concl} we summarize the results of our analysis and
present the conclusions.\\*
Throughout this work we use, depending on the context, a SCDM ($\Omega_{\mathrm{m}}=1.0$,
$\Omega_{\mathrm{\Lambda}}=0.0$) cosmology with $H_{\mathrm{0}} =
50$~km\,s$^{-1}$\,Mpc$^{-1}$ and/or a $\Lambda$CDM ($\Omega_{\mathrm{m}}=0.3$,
$\Omega_{\mathrm{\Lambda}}=0.7$) cosmology with $H_{\mathrm{0}} =
70$~km\,s$^{-1}$\,Mpc$^{-1}$. These will be denoted SCDM50 and
$\Lambda$CDM70, respectively. The main reason for presenting results for both cosmologies is that in the literature often results for the SCDM50 cosmology are listed, while only recently more people start using the $\Lambda$CDM70 model. This makes a comparison of our results with other work more easy.  

\section{Sample\label{sect:sample}}
The sample consists of 17 clusters with cooling flows listed in
 Table~\ref{tab:fit}. \cite{kaastra03} derived deprojected (3D)
 and projected (2D) radial temperature profiles for all the clusters in
 this sample. From the analysis of deprojected spectra \cite{kaastra03} give the results of single-temperature and
 multi-temperature fittings. Even though the single-temperature model
 does not provide the best fits to the inner cooling parts in several
 clusters, we will use the results from the single-phase model for the
 following reason. The multi-temperature fitting indicates that the
 gas in the cooling flow region is multiphase and at the same time
 that the emission measure distribution is dominated by the emission
 from the hottest gas at each radius (see \cite{kaastra03}, Table
 7). Using the results of the multi-temperature fitting and assuming
 pressure equilibrium between the gas phases we compute filling factors and total gas density
 for each shell. We then compute the effective temperature of the
 gas mixture in each shell and find that, for all clusters and all
 shells, it agrees within less then a few percents with the shell
 temperature from the single-temperature modelling.\\*
Spectral fitting is done over the full 0.2--10~keV range. The energy range
below 0.2~keV is currently too poorly calibrated to be useful for spectral
analysis, while above 10~keV most of our spectra lack sufficient
 flux. In general, temperatures and gas densities are computed for the innermost 8
 shells (see \cite{kaastra03}, Table 3 for the boundaries
 between the shells), except when the data become too noisy in the outermost shell.\\*
In the derivation of the total gravitational mass from the 3D temperature and density profiles and
 for the study of temperature gradients at large radii, it is crucial
 to select bins with a robust temperature estimate. For Virgo
 and Perseus the innermost shell is excised because of the presence of
 a strong X-ray emitting AGN. At high energies, the background as measured with XMM-Newton is dominated by a
 power law contribution caused by soft protons. This component is time variable
 and we estimate that we can determine it with a maximum uncertainty
 of $10 \%$ for each observation. As in the outer regions of the cluster the X-ray flux
 drops rapidly, this soft proton component dominates (see
 \cite{kaastra03}, Sect. 2.3 for a detailed presentation on the
 background subtraction). For this reason we
 exclude temperature bins in the outermost parts of the cluster as follows. Considering that the background uncertainty is of the order of $10 \%$, a bin with temperature $k T$ is selected if $10 \%$ of the net
 flux is larger or equal to $10 \%$ of the background in the energy
 band containing the turnover region of the bremsstrahlung spectrum,
 i.e. the one closer to the energy interval $k T$ - 2$ k
 T$. Concerning the choice of the energy band in the selection procedure, we notice that
 particular carefulness is needed in the selection of the outermost
 bins with low temperatures. We have applied our selection criterion using different energy bands close to the energy interval $k
 T$ - 2$ k T$ and find that if flux at too low energy compared to $kT$
 is taken into account, bins that would otherwise be excluded, satisfy
 the selection criterion. This effect can be attributed to the soft
 excess, since it is affecting the outermost bins only and it is
 strong in clusters known to show soft excess (A~2052,
 S\'ersic~159$-$3, MKW 9 and MKW 3s). A detailed modelling of
 the soft excess is beyond the scope of this work and we prefer to
 exclude bins without robust temperature estimates even though their
 inclusion would allow us to trace temperature profiles at much larger radii.
The selection is performed using both pn and MOS1/2 data.
We find the selection with pn to be more conservative than the one
 performed on MOS1/2 data and therefore use the results of the former. 
In Table~\ref{tab:fit} we list $R_{\mathrm{out}}$, the value of
 the radius at the outer end of the last radial bin considered after
 the selection. The inputs of our study are therefore: gas bolometric
 luminosity, gas density and temperature in each shell within
 $R_{\mathrm{out}}$.

\section{Gravitational masses\label{sect:masses}}
In order to make a fair comparison between clusters, the various
quantities such as gas temperature and entropy must be evaluated at constant fractions of the virial radius, rather than at fixed radii imposed by data limits. We therefore estimate the total mass assuming spherical symmetry and hydrostatic equilibrium and compute $r_{\mathrm{\Delta}}$, the radius within which the mean interior density is $\Delta$ times the critical value.\\*
We estimate the total gravitational mass by making direct use of the 3D gas temperature and gas density from the single
phase analysis. We invert the equation of hydrostatic equilibrium, i.e.
\begin{equation}\label{MDM}
M_{\mathrm{tot}}(<r)=-\frac{k T(r)}{\mu m_{\mathrm{p}} G}\Big(\frac{d\, \mathrm{ln} \rho(r)}{d\,\mathrm{ln} r}+\frac{d \, \mathrm{ln} T(r)}{d \, \mathrm{ln} r}\Big)r,
\end{equation}
where $G$ is the gravitational constant, $k$ the Boltzmann constant,
$\rho$ the gas density, $T$ its temperature and $\mu m_\mathrm{p}$ is
the mean particle mass of the gas (we assume $\mu=0.61$).
Since the mass contribution from galaxies is small we neglect it, and
therefore
$M_{\mathrm{tot}}(<r)=M_{\mathrm{DM}}(<r)+M_{\mathrm{gas}}(<r)$,
i.e. the total gravitational mass within a sphere of radius $r$ is
given by gas plus dark matter mass. 
Using the three-dimensional gas density, we select the dark matter
mass model that reproduces better the deprojected temperature
profile. As a cosmologically motivated dark matter mass model, we
consider the integrated NFW (\cite{nfw96}) dark matter profile:
\begin{equation}\label{NFW}
M_{\mathrm{DM}}(<r)=4 \pi r_{\mathrm{s}}^3 \rho_{\mathrm{c}} \frac{200}{3} \frac{c^3\Big( \mathrm{ln}(1+r/r_{\mathrm{s}}) -\frac{r/r_{\mathrm{s}}}{(1+r/r_{\mathrm{s}})}\Big)}{\mathrm{ln}(1+c)-c/(1+c)},
\end{equation}
where $\rho_{\mathrm{c}}$ is the critical density. The scale radius
$r_{\mathrm{s}}$ and the concentration parameter $c$ are the free
parameters.\\*
We make use of the NFW dark matter density profile because the choice
of an analytical function allows us to extrapolate the dark matter mass
profile beyond $R_{\mathrm{out}}$.  
The best fit parameters that minimize the $\chi^2$ of the comparison
between the temperatures predicted by Eq.(\ref{MDM}) and the observed
temperature profile are searched in the intervals $50 \, \mathrm{kpc} <
r_{\mathrm{s}} <  1.5 \, \mathrm{Mpc}$ and $2 < c < 13$ for both SCDM50 and
$\Lambda$CDM70 cosmologies.
The results of the procedure are listed in Table~\ref{tab:fit}.
We have also performed the same fitting procedure by neglecting the gas mass
contribution to the total mass and find that in this case the fitting procedure
yields substantially larger $\chi^2$.   
We do not obtain any $\chi^2$ solution for the Virgo cluster, MKW~9, Hydra~A
and A~399, which are therefore excluded from the sample. For Virgo we
obtain an unphysically small scale radius (and large concentration),
because of the small field of view. The failure of the fit
for the other clusters can be attributed to the low degree of
smoothness of their temperature profiles (no smoothing of the
data is used in our procedure).\\*
From the best fit parameters we compute $r_{\mathrm{\Delta}}$ using
\begin{equation}\label{rdelta}
\Delta=\frac{3 M_{\mathrm{tot}}(<r_{\Delta})}{4 \pi \rho_{c,z}r_{\Delta}^3 },
\end{equation} 
where $ \rho_{c,z}=(3 H_z^2)/(8 \pi G)$ is the critical density.
We estimate $r_{\mathrm{\Delta}}$ for various overdensities: $\Delta =
2500, 2000, 1500, 1000, 500, 200$ for both cosmologies, and at $\Delta
=\Delta_\mathrm{c}$, defined such that
$r_{\mathrm{\Delta_\mathrm{c}}}$ corresponds to the
cluster virial radius $r_{\mathrm{vir}}$.
For a SCDM cosmology, $\Delta_\mathrm{c}=178$, while
$\Delta_\mathrm{c}=178+82 x -39 x^2$ for $\Lambda$CDM, where
$x=\Omega(z)-1$ and $\Omega(z)=\Omega_{\mathrm{m}} (1+z)^3/(\Omega_{\mathrm{m}} (1+z)^3
+\Omega_{\mathrm{\Lambda}})$ (\cite{bryanandnorman}).   
In the computation of $r_{\mathrm{\Delta}}$ we use either 1)
$M_{\mathrm{tot}}=M_{\mathrm{DM}}+M_{\mathrm{gas}}$ or, neglecting the
gas mass contribution, 2) $M_{\mathrm{tot}}=M_{\mathrm{DM}}$.   
Assumption 1) naturally leads to values for $r_{\mathrm{\Delta}}$ that
are larger than those computed assuming 2). For both cosmologies, the relative difference
between these values is small: 4-8 percent for $\Delta=2500$ and
10-15 percent at $\Delta=500$, depending on the gas mass fraction of
the cluster. Even though assumption 1) is consistent with our
procedure for the dark matter profile estimation, we use the
$r_{\mathrm{\Delta}}$ values computed considering dark matter only for the
following reason. When gas is taken into account, one needs
to extrapolate $M_{\mathrm{gas}}(<r)$ for $r > R_{\mathrm{out}}$ when
needed. We extrapolate using a least squares fit with an integrated
$\beta$-profile with the parameter $\beta$ fixed to values taken from the
literature: for Perseus (A~426) from \cite{ettori98}, for A~3112
from \cite{sanderson03} and from \cite{peterson03} for the remaining clusters. The mean relative error measured in the
observed region is propagated to the extrapolated values. We
find that for most of the clusters in our sample extrapolation is
needed even for $\Delta=1500$ and that therefore the procedure is very
unreliable at large radii. In addition, the extrapolated values for
the gas mass are very sensitive to the value of the parameter $\beta$. 
We therefore prefer, as widely done in the analysis of numerical
simulations and in studies similar to our, assumption 2), i.e. we
neglect the gas mass contribution when computing $r_{\mathrm{\Delta}}$.\\*     
The size-temperature relation $r_{\mathrm{\Delta}} \propto
\sqrt{<T_{\mathrm{X}}>}$ predicted by self-similarity (\cite{mohr97}) and its normalization provided by catalogues of simulated
clusters allows an estimate of $r_{\mathrm{\Delta}}$ from the mean
emission-weighted temperature $<T_{\mathrm{X}}>$ alone. We 
employ the widely used predictions given by the simulation of \cite{evrard96} and compute the characteristic radii
$r_{\mathrm{\Delta}}^{\mathrm{Ev}}$. We estimate the mean
emission-weighted cluster temperature $<T_{\mathrm{X}}>$ by excluding the
cooling flow region (i.e. we excise bins with cooling time less
or equal to 15 Gyr) and fitting the remaining emission-weighted
temperature bins with a constant. We list the mean emission-weighted temperatures $<T_{\mathrm{X}}>$ in Table~\ref{tab:clusmasses}.
For both SCDM50 and $\Lambda$CDM70, we find that $r_{\mathrm{\Delta}}^{\mathrm{Ev}}$ is systematically
larger than the values we compute by directly using the best fit NFW
profiles. This is independent on whether our estimate of $r_{\mathrm{\Delta}}$ is smaller or
larger than $R_{\mathrm{out}}$ and we find that the relative difference can be as
high as 60-80 percent for $\Delta=200$. This result has also been
found by \cite{sanderson03} with a different method to determine the total
gravitational mass.\\*
From the set of $(c,r_{\mathrm{s}})$ parameters acceptable at 1
$\sigma$ we compute, for each radius $r_{\mathrm{\Delta}}$, the maximum
and minimum value for the total mass within $r_{\mathrm{\Delta}}$. In
Tables~\ref{tab:clusmasses} and~\ref{tab:clusmasseslambda} we list
$r_{\mathrm{\Delta}}$, dark matter mass and gas mass within $r_{\mathrm{\Delta}}$ for $\Delta=2500$ and
$500$ for SCDM50 and $\Lambda$CDM70, respectively. The quoted error related to the mass estimate is the root mean
square of the difference between maximum and mean, and the difference
between mean and minimum. When extrapolation is needed, the values for the gas mass are
computed using the least squares fit described above. We compared our
estimates with results from the literature. For clusters with
published quantities comparable to ours, we generally find good
agreement:
\begin{itemize}
\item A~262: \cite{fino}, using temperature profiles derived
with ASCA, estimate $r_{500}=860 \pm 170$ kpc and
$M_{\mathrm{tot}}(<r_{500})=9.3 \pm 2.7 \times 10^{13}M_{\odot}$ (in SCDM50). Our estimate of $r_{500}$ (see Table \ref{tab:clusmasses}) is lower
probably due to the fact that contrary to \cite{fino} we do not
include the gas mass contribution in its computation. Consequently
also $M_{\mathrm{tot}}(<r_{\mathrm{500}})$ is lower, but when computing
$M_{\mathrm{tot}}(<860 $\, kpc$)=6.7 \pm 1.8 \times 10^{13}M_{\odot}$ we
find consistency for the total mass estimate.      
\item 2A~0335+096: as for A~262 the estimate for $r_{500}$ of \cite{fino} is at
variance with our, but agreement is found when the total mass is
compared at the same radius: $34.4 \pm 11.3 \times 10^{13}M_{\odot}$ and $26.1
\pm 3.64 \times 10^{13}M_{\odot}$ being \cite{fino}'s and our estimate,
respectively, at 1340 kpc. \cite{ettori02b} derived total and gas
masses from {\it BeppoSAX} data at $\Delta=2500$ and 1000 using a
method similar to our, with the only difference being that their NFW
profile estimates dark matter and gas masses altogether and not dark
matter only as in our analysis. Due to this difference and to the
larger radial range they access ($R_{\mathrm{out}}=917$ kpc with
respect to $R_{\mathrm{out}}=510$ kpc in our work), the best fit scale
radius is larger ($r_{\mathrm{s}}=626$ kpc) than our estimate and
consequently the concentration parameter smaller. The characteristic
radii $r_{\Delta}$ estimated by \cite{ettori02b} are also larger than
ours, again due to the fact that we neglect the gas mass in the computation. Nevertheless, we find good agreement between the total
and gas masses when evaluated at the same physical radius
rather than at the same overdensity.      
\item MKW~3s: we find excellent agreement between the estimates of \cite{fino}, $r_{500}=1210 \pm 140$ kpc and
$M_{\mathrm{tot}}(<r_{500})=25.7 \pm 4.3 \times 10^{13}M_{\odot}$, and
ours (see Table \ref{tab:clusmasses} for our values of
$M_{\mathrm{DM}}$ and $M_{\mathrm{gas}}$). 
\item A~2052: we find excellent agreement between the estimates of \cite{fino}, $r_{500}=1280 \pm 280$ kpc and
$M_{\mathrm{tot}}(<r_{500})=30.2 \pm 9.6 \times 10^{13}M_{\odot}$ and
ours (see Table \ref{tab:clusmasses}).  
\item A~4059: the values given by \cite{fino} ($r_{500}=1390 \pm 300$ kpc and
$M_{\mathrm{tot}}(<r_{500})=38.7 \pm 12.0 \times 10^{13}M_{\odot}$)
are in agreement with ours (see Table \ref{tab:clusmasses}).  
\item A~496: as for A~262 and 2A~0335+096 the estimate of $r_{500}$ of \cite{fino} is at
variance with our. Disagreement is still found for the total mass estimated for
$r_{500}=1480$ kpc: $47.0 \pm 6.7 \times 10^{13}M_{\odot}$ and $19.9
\pm 9.7 \times 10^{13}M_{\odot}$ are \cite{fino}'s and our estimate,
respectively. As for 2A~0335+096 our estimates can also be compared to
\cite{ettori02b}'s results. We find discrepancy as in the case of
2A~0335+096: due to the much larger radial range they access ($R_{\mathrm{out}}=845$ kpc with
respect to $R_{\mathrm{out}}=319$ kpc in our work), the best fit NFW scale
radius is much larger ($r_{\mathrm{s}}=738$ kpc) than our estimate
(see Table \ref{tab:fit}) and consequently the concentration parameter
smaller. Furthermore, even when compared at the same radius, our results for both
total and gas masses are $\approx 50 \%$ smaller than
\cite{ettori02b}'s. The disagreement might be due to the fact that the
comparison is made at radii much larger than $R_{\mathrm{out}}$, where
the mass extrapolation is large.
\item A~3112: we find excellent agreement between the estimates of \cite{fino}, $r_{500}=1350 \pm 120$ kpc and
$M_{\mathrm{tot}}(<r_{500})=35.4 \pm 4.5 \times 10^{13}M_{\odot}$ and
ours (see Table \ref{tab:clusmasses}).
\item A~1795: agreement is found between the estimates of \cite{fino}, $r_{500}=2000 \pm 560$ kpc and
$M_{\mathrm{tot}}(<r_{500})=116.6 \pm 47.2 \times 10^{13}M_{\odot}$ and
ours (see Table \ref{tab:clusmasses}). Better agreement is found when
masses are evaluated at the same radius: we estimate
$M_{\mathrm{tot}}(<2000$\,kpc$)=82.8 \pm 10.2 \times
10^{13}M_{\odot}$. As for 2A~0335+096 and A~496 we find a difference
between our best fit NFW parameters and those computed by
\cite{ettori02b} (they can use $R_{\mathrm{out}}=1584$ kpc with
respect to $R_{\mathrm{out}}=900$ kpc in this work). The difference
between \cite{ettori02b}'s characteristic radii and total and gas
masses is nevertheless small and good agreement is found when these
quantities are computed at the same radii,
i.e. $M_{\mathrm{tot}}(<771$\, kpc$)=40.0 \pm 1.9 \times
10^{13}M_{\odot}$ and $M_{\mathrm{gas}}(<771$\, kpc$)=53.7 \pm 0.7 \times
10^{12}M_{\odot}$ from \cite{ettori02b} versus
$M_{\mathrm{tot}}(<771$\, kpc$)=37.5 \pm 2.5 \times
10^{13}M_{\odot}$ and $M_{\mathrm{gas}}(<771$\, kpc$)=47.7 \pm 0.8 \times
10^{12}M_{\odot}$ in this work.              
\item A~426: even though there is a difference between the NFW
parameters estimated by \cite{ettori02b} and ours we find excellent
agreement between the values of the characteristic radii and both
total and gas masses.  
\item A~1835: \cite{allen} derived total masses from {\it Chandra} data at $\Delta=2500$ for both SCDM50 and
$\Lambda$CDM70 and using a method similar to our. For both cosmologies,
we find good agreement with their best fit NFW profile
parameters and estimated $r_{2500}$. The small differences, in particular the
smaller values we obtain for $r_{2500}$, are very likely due to the
fact that in their method the gas mass contribution is also included
in the computation of $r_{2500}$. When total masses are compared at
the same radii, good agreement is found, i.e., for SCDM50,
$M_{\mathrm{tot}}(<720$\, kpc$)=5.41^{+1.13}_{-0.76} \times
10^{14}M_{\odot}$ and $M_{\mathrm{tot}}(<720$\, kpc$)=4.56 \pm  0.35 \times
10^{14}M_{\odot}$ being \cite{allen}'s value and our estimate,
respectively.
\item For NGC~533, A~1837 and S\'ersic~159$-$3 a fair comparison with
previous work is not possible.   
\end{itemize}
To reiterate, the remainder of our analysis of our final sample of 13
objects relies on the validity of
extrapolating profiles fitted to the inner region of the data: for
both SCDM50 and $\Lambda$CDM70, the
number of objects for which the outer radius $R_{\mathrm{out}}$ is
larger than (or equal to) $r_{\mathrm{\Delta}}$ at the 95 percent
level of confidence is: 11 (NGC~533, A~262, A~1837, S\'ersic~159-3,
2A~0335+096, MKW~3s, A~2052, A~4059, A~3112, A~1795 and A~1835) for
$\Delta=2500$, 7 (NGC~533, A~1837, S\'ersic~159-3, MKW~3s,
A~4059, A~1795 and A~1835) for $\Delta=1500$, only 3 (A~1837,
S\'ersic~159-3 and A~1835) for $\Delta=1000$ and none for
$\Delta=500$.

\begin{table*}[!ht]
\caption{\label{tab:fit} The radius at the outer end of the last radial bin
considered, $R_{\mathrm{out}}$ and the best fit parameters for the NFW
dark matter density profile for SCDMH50 and $\Lambda$CDMH70. Virgo, Hydra~A and A~399 do not
converge in the $r_{\mathrm s}-c$ plane. The quoted errors on $r_{\mathrm{s}}$ and $c$ are RMS of
the 68 percent joint confidence limits.}
\centerline{
\begin{tabular}{|l|rrrr|rrrr|}
\hline
\multicolumn{1}{|c|}{} &
\multicolumn{4}{|c|}{SCDMH50} &
\multicolumn{4}{|c|}{$\Lambda$CDMH70}\\
 Cluster& $R_{\mathrm{out}}$& $r_{\mathrm{s}}$&$c$ &$\chi^2 \, \mathrm{(d.o.f.)}$& $R_{\mathrm{out}}$ &$r_{\mathrm{s}}$&$c$ &$\chi^2 \, \mathrm{(d.o.f.)}$\\
 &(kpc)&(kpc)& & &(kpc)&(kpc)& & \\
\hline
NGC 533               &267&52(4)&12.56(0.56)&3.68(5)&192&37(3)&12.53(0.55)&3.78(5)\\
Virgo                 &56&-&-&-&40&-&-&-\\
A 262                 &316&110(22)&9.03(1.03)&8.07(6)&228&85(17)&8.62(1.00)&7.73(6)\\
A 1837                &656&275(117)&6.07(2.51)&1.07(3)&485&211(90)&5.78(2.66)&1.08(3)\\       
S\'ersic~159$-$3      &814&209(25)&6.79(0.57)&54.36(5)&598&159(18)&6.56(0.54)&53.13(5)\\
MKW 9                 &262&-&-&-&191&-&-&-\\
2A 0335+096           &510&228(21)&6.45(0.35)&8.06(5)&371&203(20)&5.63(0.33)&13.94(5)\\
MKW 3s                &661&280(30)&6.19(0.45)&26.60(5)&483&213(23)&5.96(0.43)&26.39(5)\\
A 2052                &526&311(56)&5.40(0.59)&73.47(5)&383&260(48)&4.90(0.55)&87.73(5)\\
A 4059                &675&1002(404)&2.76(0.99)&1.92(4)&494&744(297)&2.70(1.00)&1.91(4)\\
Hydra A  (A 780)      &349&-&-&-&256&-&-&-\\
A 496                 &319&166(64)&8.11(1.72)&0.46(4)&232&129(51)&7.75(1.67)&0.45(4)\\
A 3112                &464&462(112)&4.78(2.00)&1.97(3)&344&361(166)&4.54(2.04)&1.82(3)\\
A 1795                &900&507(57)&4.70(0.32)&14.37(5)&664&430(49)&4.21(0.30)&9.04(5)\\
A 399                 &983&-&-&-&727&-&-&-\\
Perseus (A 426)       &363&629(248)&4.21(0.79)&11.56(4)&261&481(189)&4.09(0.77)&12.30(4)\\
A 1835                &1191&799(159)&3.67(0.44)&5.32(3)&950&610(150)&3.39(0.41)&5.30(3)\\
\hline
\end{tabular}
}
\end{table*}

\begin{table*}[!ht]
\caption{\label{tab:clusmasses} Mean emission-weighted temperature $<T_{\mathrm{X}}>$ and
characteristic radii $r_{\mathrm{\Delta}}$, dark matter mass
$M_\mathrm{DM}(<r_\mathrm{\Delta})$ and gas
mass $M_{\mathrm{gas}}(<r_\mathrm{\Delta})$ for overdensities $\Delta=2500$ and
$500$ in the SCDM50 cosmology.}
\centerline{
\begin{tabular}{|lrrrrrrr|}
\hline
Cluster
&$<T_{\mathrm{X}}>$&$r_\mathrm{2500}$&$M_\mathrm{DM}(<r_\mathrm{2500})$&$M_{\mathrm{gas}}(<r_\mathrm{2500})$&$r_\mathrm{500}
  $&$M_\mathrm{DM}(<r_\mathrm{500})$&$M_{\mathrm{gas}}(<r_\mathrm{500})$ \\
 &(keV)&(kpc)&($10^{13}M_{\odot}$)&($10^{12}M_{\odot}$)&(kpc)&($10^{13}M_{\odot}$)&($10^{12}M_{\odot}$)  \\
\hline
NGC 533            &1.21(0.05)&220(5)&0.81(0.04)&0.56(0.08)&439(12)&1.30(0.09)&1.89(0.27)\\
A 262              &2.17(0.01)&322(21)&2.53(0.34)&2.95(0.42)&663(51)&4.44(0.80)&9.24(1.46)\\
A 1837             &4.00(0.13)&466(51)&9.02(1.77)&9.03(2.16)&1021(208)&18.99(8.37)&29.08(9.50)\\
S\'ersic~159$-$3   &2.29(0.07)&413(9)&6.05(0.25)&10.53(0.60)&888(28)&12.02(0.86)&25.17(1.97)\\
2A 0335+096        &2.94(0.05)&439(12)&6.79(0.35)&11.69(1.13)&947(32)&13.63(1.03)&26.02(2.79)\\
MKW 3s             &3.48(0.14)&503(11)&10.59(0.44)&13.59(0.76)&1095(34)&21.80(1.51)&35.28(2.27)\\
A 2052             &2.99(0.08)&481(21)&9.00(0.74)&11.33(1.00)&1068(65)&19.63(2.60)&31.64(3.15)\\
A 4059             &3.96(0.05)&640(75)&21.88(3.77)&18.30(3.41)&1637(372)&73.05(30.38)&49.77(12.66)\\
A 496              &4.20(0.19)&420(48)&5.93(1.37)&9.65(1.96)&880(123)&10.88(3.55)&25.39(5.53)\\
A 3112             &4.33(0.11)&578(104)&17.47(5.18)&20.98(5.15)&1321(390)&41.64(25.59)&54.49(20.59)\\
A 1795             &5.62(0.04)&631(16)&22.02(1.02)&36.29(2.09)&1443(54)&52.58(4.15)&96.49(6.31)\\
Perseus (A 426)    &5.30(0.04)&734(89)&30.32(6.41)&26.73(3.77)&1694(272)&74.55(25.13)&53.26(8.18)\\
A 1835             &6.78(0.37)&534(12)&21.78(0.77)&63.18(3.78)&1353(73)&70.93(7.26)&145.52(14.23)\\
\hline\noalign{\smallskip}  
\end{tabular}
}
\end{table*}


\begin{table*}[!ht]
\caption{\label{tab:clusmasseslambda} Characteristic radii
$r_{\mathrm{\Delta}}$, dark matter mass
$M_\mathrm{DM}(<r_\mathrm{\Delta})$ and gas
mass $M_{\mathrm{gas}}(<r_\mathrm{\Delta})$ for overdensities $\Delta=2500$ and
$500$ in the $\Lambda$CDM70 cosmology.}
\centerline{
\begin{tabular}{|lrrrrrr|}
\hline
Cluster
&$r_\mathrm{2500}$&$M_\mathrm{DM}(<r_\mathrm{2500})$&$M_{\mathrm{gas}}(<r_\mathrm{2500})$&$r_\mathrm{500}
  $&$M_\mathrm{DM}(<r_\mathrm{500})$&$M_{\mathrm{gas}}(<r_\mathrm{500})$ \\
 &(kpc)&($10^{13}M_{\odot}$)&($10^{12}M_{\odot}$)&(kpc)&($10^{13}M_{\odot}$)&($10^{12}M_{\odot}$)  \\
\hline
NGC 533            &160(4)&0.59(0.03)&0.25(0.04)&319(9)&0.94(0.06)&0.84(0.12)\\
A 262              &239(16)&1.97(0.27)&1.36(0.20)&495(39)&3.50(0.64)&4.29(0.69)\\
A 1837             &362(42)&7.24(1.51)&4.63(1.14)&793(165)&15.17(6.88)&14.59(4.82)\\
S\'ersic~159$-$3   &320(7)&4.94(0.21)&5.24(0.31)&688(22)&9.77(0.69)&12.28(0.96)\\
2A 0335+096        &344(10)&5.98(0.33)&5.75(0.57)&754(28)&12.60(1.02)&12.73(1.40)\\
MKW 3s             &384(8)&8.42(0.36)&6.59(0.37)&835(26)&17.31(1.21)&16.90(1.09)\\
A 2052             &372(17)&7.56(0.65)&5.56(0.50)&833(54)&17.02(2.37)&15.51(1.60)\\
A 4059             &491(62)&17.6(3.25)&8.88(1.73)&1247(304)&57.72(25.74)&23.65(6.31)\\
A 496              &319(37)&4.74(1.12)&4.61(0.95)&668(96)&8.76(2.94)&12.06(2.68)\\
A 3112             &460(86)&14.86(4.63)&10.78(2.74)&1048(317)&35.16(22.24)&27.85(6.40)\\
A 1795             &501(13)&19.05(0.88)&18.71(1.07)&1157(45)&46.87(3.77)&49.18(3.25)\\
Perseus (A 426)    &555(68)&24.69(5.22)&12.29(1.72)&1282(207)&61.03(20.59)&24.32(3.72)\\
A 1835             &477(43)&19.93(2.86)&40.46(5.28)&1175(60)&59.61(26.48)&88.05(20.44)\\
\hline\noalign{\smallskip}  
\end{tabular}
}
\end{table*}


\section{Scaled temperature profiles\label{sect:Tprofiles}}
\begin{figure*}
\resizebox{\hsize}{!}{\includegraphics{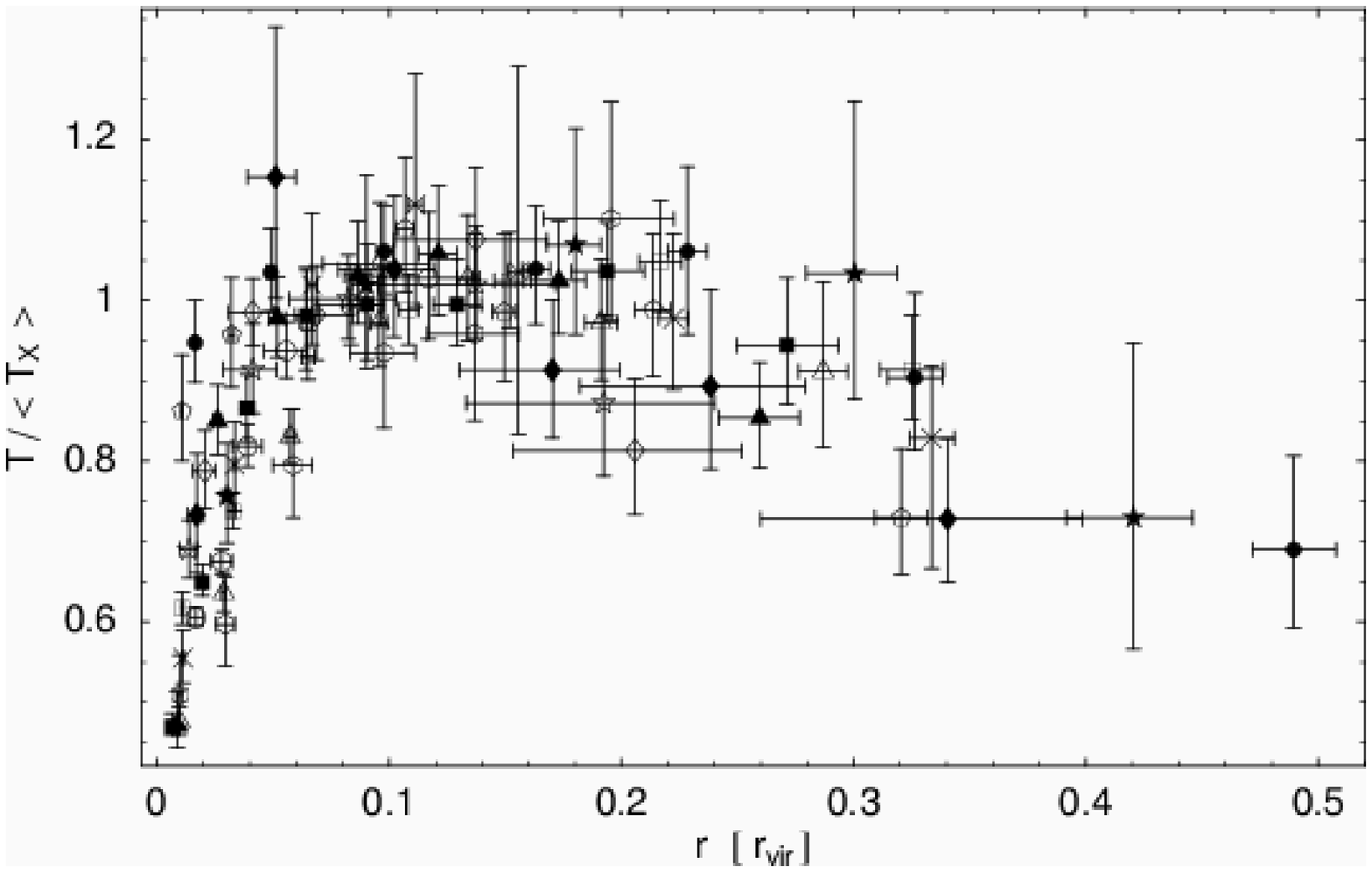}}
\caption{Scaled temperature profiles (deprojected) in SCDM50 cosmology: the radius is
scaled with the virial radius $r_{\mathrm{vir}}=r_{\mathrm{180}}$, while the temperature is normalized by the mean
emission-weighted temperature $<T_{\mathrm{X}}>$. Clusters are related
to symbols as follows: NGC 533 ({\sl crosses}), A 262 ({\sl filled
squares}), A 1837 ({\sl filled diamonds}), S\'ersic~159$-$3 ({\sl filled
circles}), 2A 0335+096 ({\sl open triangles}),
MKW 3s ({\sl open pentagons}), A 2052 ({\sl filled triangles}), A 4059
({\sl open diamonds}), A 496 ({\sl open hexagon}), A 3112 ({\sl open
stars}), A 1795 ({\sl open squares}), Perseus ({\sl open circles}) and A
1835 ({\sl filled stars}).}
\label{fig:Tscaled}
\end{figure*}

\begin{figure*}
\resizebox{\hsize}{!}{\includegraphics{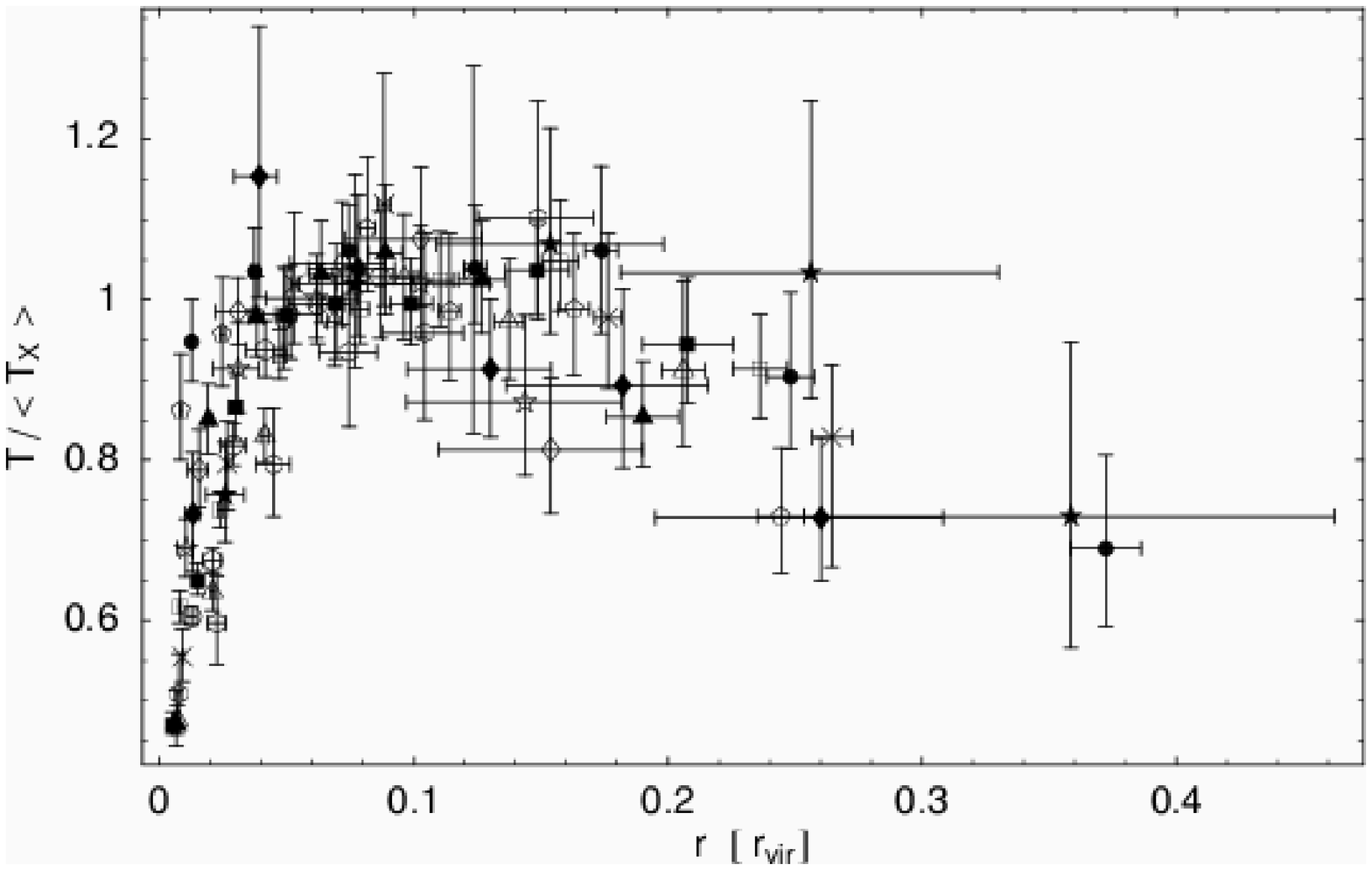}}
\caption{Same as Figure \ref{fig:Tscaled}, but in $\Lambda$CDM70
cosmology.}
\label{fig:Tscaledlambda}
\end{figure*}

In the following we investigate the structure of the temperature
profiles for our final sample of 13 cooling flow clusters. In
Figs.~\ref{fig:Tscaled} (for SCDM50) and \ref{fig:Tscaledlambda} (for $\Lambda$CDM70), we present the deprojected radial profiles plotted
against the radius in units of $r_{\mathrm{vir}}$ ($\approx
r_{\mathrm{180}}$ for SCDM and $\approx
r_{\mathrm{101}}$ for $\Lambda$CDM at $z=0$), where the temperature has been
normalized by the mean emission-weighted temperature
$<T_{\mathrm{X}}>$. From a visual inspection it is evident that a
temperature gradient is present at large radii and that when
normalized and scaled by the virial radius, temperature profiles are
remarkably similar. In addition almost all the individual profiles clearly show a break radius $r_{\mathrm{br}}$, a
decrease of temperature from $r_{\mathrm{br}}$ inwards and a decline
at radii larger than $r_{\mathrm{br}}$. In the following, we
investigate the shape in detail.

\subsection{The break radius\label{sect:break}}
\begin{figure*}
\resizebox{\hsize}{!}{\includegraphics{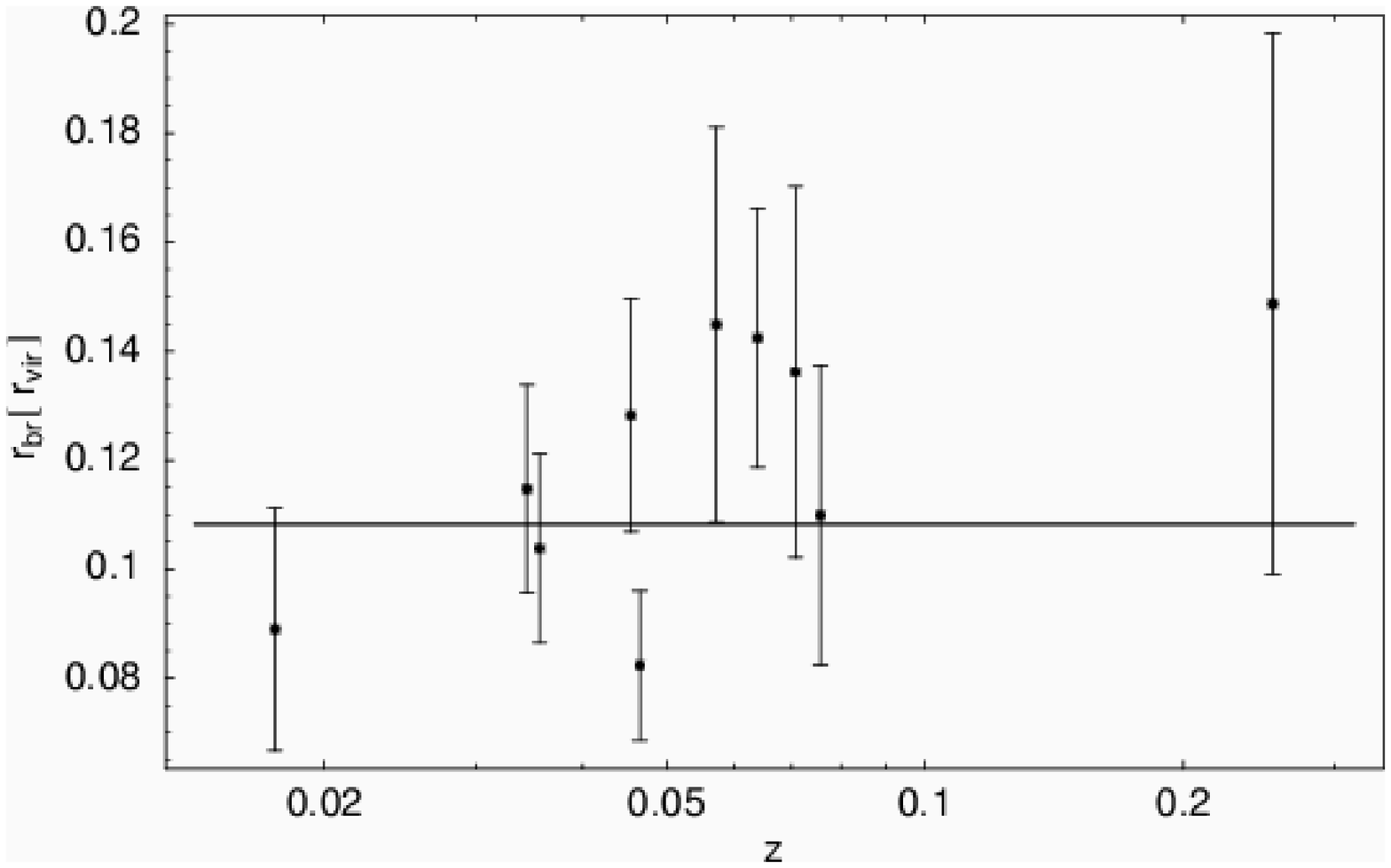}}
\caption{Break radii $r^{\mathrm{proj}}_{\mathrm{br}}$ in units of $
r_{\mathrm{vir}}$ as a function
of redshift in SCDM50. The horizontal line is
the error-weighted mean value.}    
\label{fig:breakradii}
\end{figure*}

In the following we compute the break radius of the individual
clusters in units of $r_{\mathrm{vir}}$. For each cluster we divide the scaled
temperature profiles in two radial intervals: from the innermost bin to the bin $i$ and from
the bin $i+1$ to the outermost one.
Temperature profiles in each of the two intervals are then fitted
using straight lines, power laws and exponential functions. All the
nine combinations are used and for each pair of
fitting functions, $i$ is varied until the best fit is achieved. Clusters that do not show a clear temperature decrease in the outer
region are excluded. These are A~262, A~496 and Perseus. For the remaining clusters the bin $i$ which gives the
best fit is independent of the choice of the fitting functions. The break radius
$x_{\mathrm{br}}=r_{\mathrm{br}} /r_{\mathrm{vir}}$ is then defined as
$(x_{\mathrm{i}}+x_{\mathrm{i+1}})/2$ (where $x_{\mathrm{i}}$ is the
distance of the bin $i$ from the center in units of
$r_{\mathrm{vir}}$) and its uncertainty
$(x_{\mathrm{i+1}}-x_{\mathrm{i}})/2$.
In Fig.~\ref{fig:breakradii} we show the break radius
$x^{\mathrm{proj}}_{\mathrm{br}}=r^{\mathrm{proj}}_{\mathrm{br}} /r_{\mathrm{vir}}$ of
the {\sl projected} temperature profiles in SCDM50 as a function of redshift. The mean value of 
$x^{\mathrm{proj}}_{\mathrm{br}}$ is 0.11 with a standard deviation
of 0.01. Using a different method, \cite{dm02} find 0.20 for the mean value of
$x^{\mathrm{proj}}_{\mathrm{br}}$ for a sample of 11 cooling flow and
10 non-cooling flow clusters and a lower value, 0.16, for the cooling
flow clusters only. Taking into account that \cite{dm02}, who also
used {\sl projected} profiles, computed the break
radius for cooling flow clusters by excluding the cooling region and
fitting the profiles with a constant temperature for
$r<r^{\mathrm{proj}}_{\mathrm{br}}$ and with a line for $r>r^{\mathrm{proj}}_{\mathrm{br}}$, the
fact that their estimate is larger than ours is not surprising. In
$\Lambda$CDM70 the mean value of 
$x^{\mathrm{proj}}_{\mathrm{br}}$ is reduced to 0.08 with a standard deviation
of 0.01.       
In agreement with \cite{dm02} we find that the intrinsic
dispersion of the parent population of scaled break radii (assumed to
be distributed like a Gaussian) is consistent with 0 (for both SCDM50
and $\Lambda$CDM70). By performing the
same analysis using {\sl deprojected} temperature profiles we find
0.12 and 0.09  for the mean value of
$x^{\mathrm{deproj}}_{\mathrm{b}}$ for SCDM50
and $\Lambda$CDM70, respectively, and standard deviation of 0.01. This latter results shows consistency
between break radii of projected and deprojected profiles.

\subsection{The outer region\label{sect:outer}}
\begin{figure*}
\resizebox{\hsize}{!}{\includegraphics{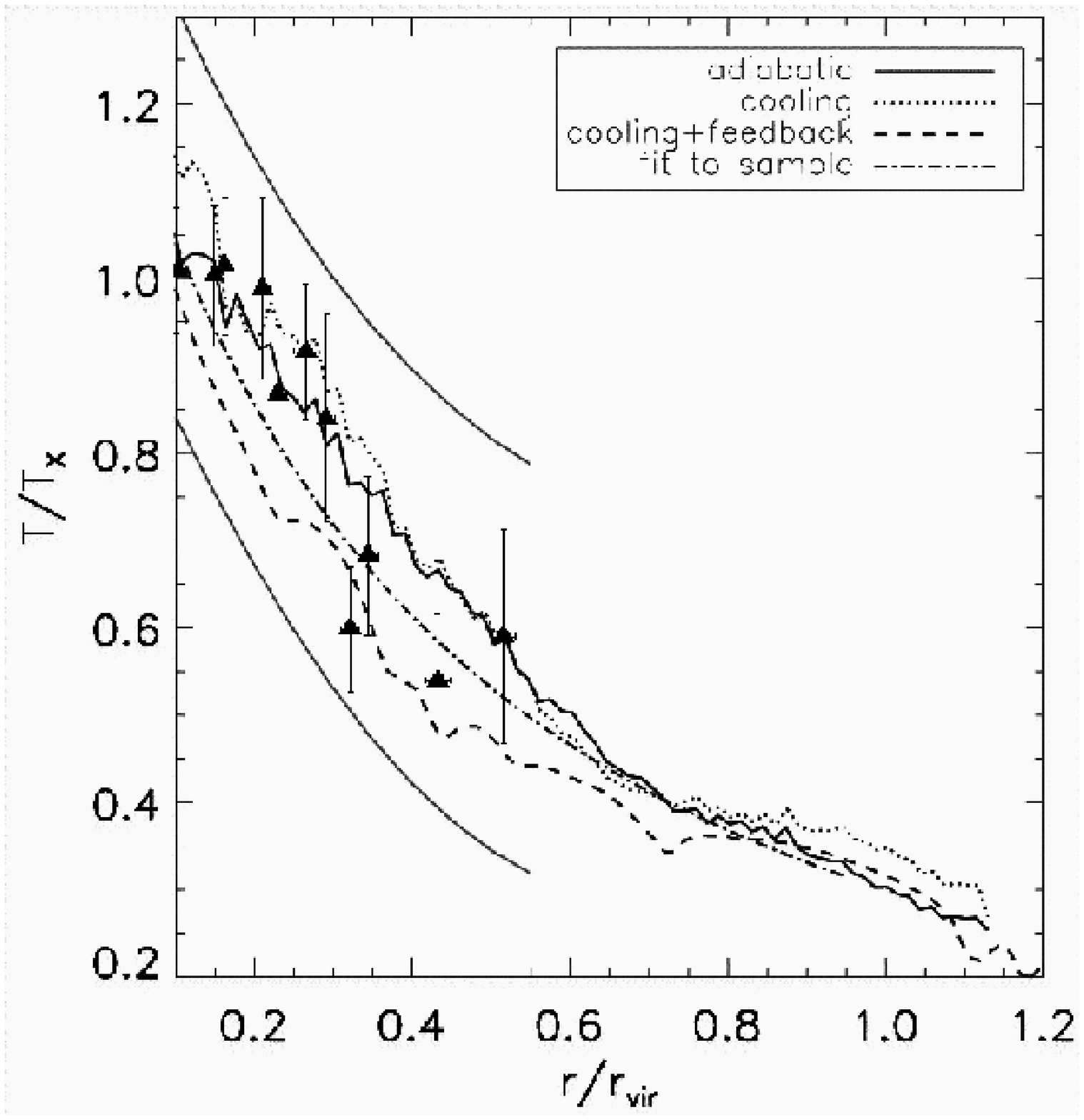}}
\caption{Mean error-weighted {\sl projected} temperature profile for
our sample (filled triangles, with 1$\sigma$ errors) as a function of radius in
units of $r_{\mathrm{vir}}$ together with the results shown
in Fig. 6 of Loken et al. (2002). The two solid lines define the approximate
band that encloses most temperature profiles and their error bars of Markevitch
et al. (1998).}
\label{fig:loken2final}
\end{figure*}

For comparison with other studies, we quantify the decline seen in
Fig.~\ref{fig:Tscaled} (i.e. in a SCDM50 cosmology) for radii larger than $\sim 0.15 r_{\mathrm{vir}}$ by simply modeling the data with a line. From the data shown in
Fig.~\ref{fig:Tscaled} we select, for each object, bins at distances
larger than $r^{\mathrm{deproj}}_{\mathrm{br}}=0.12 \,  r_{\mathrm{vir}}$.
We fit the selected points to a straight line and find a slope equal
to $-0.97 \pm 0.14$ ($\chi^2=20.38$, $\mathrm{d.o.f.}=34$).\\* 
Since many authors use {\sl projected} temperature profiles, we apply the
same procedure to the {\sl projected} temperature profiles. These are
scaled by $r_{\mathrm{180}}$ and normalized by the mean
emission-weighted temperature $<T^{\mathrm{proj}}_{\mathrm{X}}>$,
i.e. computed using the projected data. In this case we obtain
$-0.94 \pm 0.11$ ($\chi^2=26.24$, $\mathrm{d.o.f.}=35$) for the
slope.\\*
For a fair comparison with \cite{dm02}, we compute the value for the
slope using their procedure: we use {\sl projected} temperature
profiles (normalized by $<T^{\mathrm{proj}}_{\mathrm{X}}>$ and scaled
by $r_{180}^{\mathrm{Ev}}$), exclude temperature bins within the
cooling region and fit the data with the broken line model given in
Eq. (1) in \cite{dm02}, i.e. with a constant $t_b$ up to a break
radius $x_b$ and with a line with slope $m$ beyond $x_b$. In excellent
agreement with \cite{dm02}, we find $t_b=1.02 \pm 0.02$, $x_b=0.10 \pm
0.05$ and a slope $m=-1.15 \pm 0.10$. We also model the profiles scaled in the SCDM50 cosmology with a polytropic equation of state,
$T/<T_{\mathrm{X}}>=T_0 (1+(r/a_{\mathrm{x}})^2)^{-3 \beta (\gamma
-1)/2}$. We take values for $\beta$ and $a_{\mathrm{x}}$ from the
literature: for Perseus (A~426) from \cite{ettori98}, for A~3112
from \cite{sanderson03} and from \cite{peterson03} for the remaining
clusters. We then use the mean values
$\beta=0.69$, $a_{\mathrm{x}}=0.087 \, r_{\mathrm{180}}$ in the fit. For
$r > r^{\mathrm{deproj}}_{\mathrm{br}}$ we find $T_0=1.36 \pm 0.07$ and $\gamma= 1.24
\pm 0.03$, in agreement with the results of \cite{markevitch}.\\* 
Recent numerical simulation are in general able to
produce temperature profiles quite similar to the observed ones at
large radii (see \cite{dm02}, in particular Fig. 12, for a comparison
with simulations results published before 2001). Recently, \cite{dm02}'s profiles have been compared with simulated profiles of
\cite{valdarnini,borgani02} and \cite{borgani04}, and for radii larger that $\approx 0.15
r_{\mathrm{vir}}$ it was found good agreement. Given the agreement
between \cite{dm02}'s profiles and ours at large
cluster-centric distances, these
simulations should provide a good description of the temperature profiles
presented here as well. Here, we compare the {\sl projected} temperature profiles with the numerical
predictions by \cite{loken} and find good agreement. \cite{loken}
compared their results and found good agreement with observations of \cite{markevitch} and the {\it BeppoSAX}
profiles presented in \cite{dm02}. The
comparison with \cite{loken}'s results is  shown in
Fig.~\ref{fig:loken2final}, where we plot the mean error-weighted {\sl
projected} profile for $r > 0.1 r_{\mathrm{vir}}$ for our sample of
13 cooling flow clusters on top of the results shown in Fig. 6 of
\cite{loken}. \cite{loken} find
that the simulated clusters in their catalogs (both $\Lambda$CDM and
SCDM catalogs) show temperature profiles
which can be well described by a universal temperature profile of the
form $T/<T_{\mathrm{X}}>=T_0 (1+r/a_{\mathrm{x}})^{-\delta}$, with
$T_0=1.33$, $a_{\mathrm{x}}=r_{\mathrm{vir}}/1.5$ and $\delta=1.6$.
This profile is clearly not suitable to reproduce the profile
typical of cooling flow cluster over the entire observed radial range
and a central region must be excised. The
radial range used by \cite{loken} in the fit is $(0.04-1.00) \, r_{\mathrm{vir}}$. 
Due to the limited radial range of our data, we fit the universal
profile found by \cite{loken} to the {\sl projected} temperature profiles for $<r^{\mathrm{proj}}_{\mathrm{br}}/r_{\mathrm{180}}>=0.11 \leq r/r_{\mathrm{180}} \leq
 r_{\mathrm{max}}/r_{\mathrm{180}}=0.49$ by assuming
$a_{\mathrm{x}}=r_{\mathrm{180}}/1.5$. We find $T_0=1.31 \pm 0.05$ and
$\delta=1.67 \pm 0.20$. Despite the good agreement, we stress that a
more robust comparison will be only possible with observations spanning a
much larger radial range than the one used in the present work.

\subsection{The inner region}
The temperature decline towards the center is characteristic of cooling
flow clusters. For the sample in this paper, the structure of {\sl
deprojected} temperature profiles has been extensively analyzed in \cite{kaastra03}. The parametrization:  
\begin{equation}\label{eqn:Tfit}
T(r) = T_c + (T_h-T_c) \frac{(r/r_\mathrm{c})^{\mu}}
{1+(r/r_\mathrm{c})^{\mu}},
\end{equation} 
is found to provide a remarkably good description of the temperature
decline, with $T_{\mathrm{c}}$ set equal to the temperature of the
innermost bin and $\mu=2$. In SCDM50 a correlation between the fitting
parameters $T_{\mathrm{h}}$ and $r_{\mathrm{c}}$ of the form $r_c =
(4.1\pm 0.8) T_h^{1.84\pm 0.14}$, with $r_c$ in kpc and $T_h$ in keV,
is found. Assuming that $r_{\mathrm{\Delta}}$ scales with $T_h^{1/2}$
one concludes that the characteristic radius $r_c$ does not simply
scale with the virial radius.\\*
In the following we further explore this implication by considering the mean
emission-weighted temperature $<T_X>$ instead is $T_h$ and using the
estimated $r_{\mathrm{\Delta}}$ instead of assuming
$r_{\mathrm{\Delta}} \propto T_h^{1/2}$. For both SCDM50 and
$\Lambda$CDM70 cosmologies we find that the
characteristic length scale $r_c$ is not linked to the size of the dark
matted halo hosting the cooling flow in a simple way. In SCDM50, for
example, we find $r_c \propto <T_X>^{1.78\pm 0.10}$ and
$r_{\mathrm{2500}} \propto <T_X>^{0.88\pm 0.02}$, with $r_c$ and
$r_{\mathrm{2500}}$ in kpc and $<T_X>$ in keV. In this context
we notice that the flattening of the composite scaled temperature
profile in the range $\sim 0.1-0.2 r_{\mathrm{vir}}$ in SCDM50 (see
Fig.~\ref{fig:Tscaled}) is artificial,
as it is due to the fact that temperature profiles in the inner region
do not simply scale with the virial radius.\\*
\cite{kaastra03} searched for correlations between parameters of the cooling gas
without finding any significant one (\cite{kaastra03}, section 4.1). Having additional information about
the characteristic radii $r_{\mathrm{\Delta}}$, we investigated
correlations of these with the parameters of the cooling gas, such as
relative temperature decrement and cooling radius. We do not find any
significant correlation.\\*
As shown in the previous section, for radii larger than $\sim 0.15
r_{\mathrm{vir}}$ most of the numerical simulations are able to
produce temperature profiles quite similar to the ones presented in
this work. The fair comparison with simulated cluster temperature
profiles in the central region, i.e. for
radii smaller than the break radius $r_{\mathrm{br}}$, is not possible
yet. In fact, the central temperature drop has not been
reproduced consistently by any simulation (e.g.,
\cite{tornatore,borgani04} and references therein). In this context, our temperature parametrization
for the inner region of cooling flow clusters (Eq. \ref{eqn:Tfit}) and
the correlation between its free parameters ($r_c \propto
<T_X>^{1.78\pm 0.10}$) provides an important test for forthcoming
simulation models.  

\section{Entropy profiles and the L-T relation\label{sect:entropy}}
\begin{figure*}
\resizebox{\hsize}{!}{\includegraphics{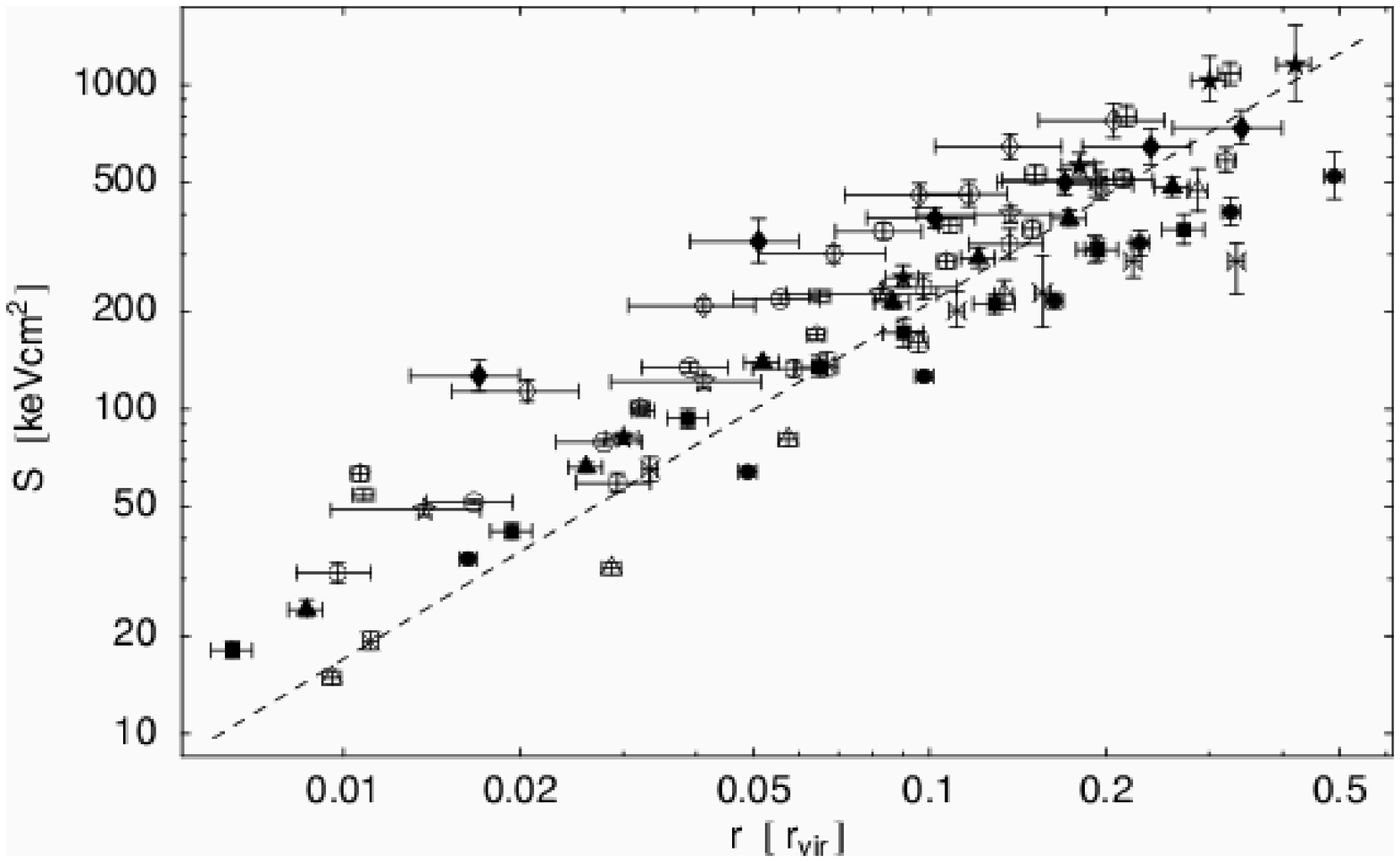}}
\caption{Entropy profiles in SCDM50: the radius is scaled with the
radius $r_{\mathrm{vir}}=r_{\mathrm{180}}$, while no scaling has been applied to
the entropy values. Clusters are related
to symbols as follows: NGC 533 ({\sl crosses}), A 262 ({\sl filled
squares}), A 1837 ({\sl filled diamonds}), S\'ersic~159$-$3 ({\sl filled
circles}), 2A 0335+096 ({\sl open triangles}),
MKW 3s ({\sl open pentagons}), A 2052 ({\sl filled triangles}), A 4059
({\sl open diamonds}), A 496 ({\sl open hexagon}), A 3112 ({\sl open
stars}), A 1795 ({\sl open squares}), Perseus ({\sl open circles}) and
A~1835 ({\sl filled stars}). The dashed line indicates the best fit
power-law with the power index value fixed to $1.1$  predicted by \cite{tozzi01}.}
\label{fig:entropynonscaled}
\end{figure*}

\begin{figure*}
\resizebox{\hsize}{!}{\includegraphics{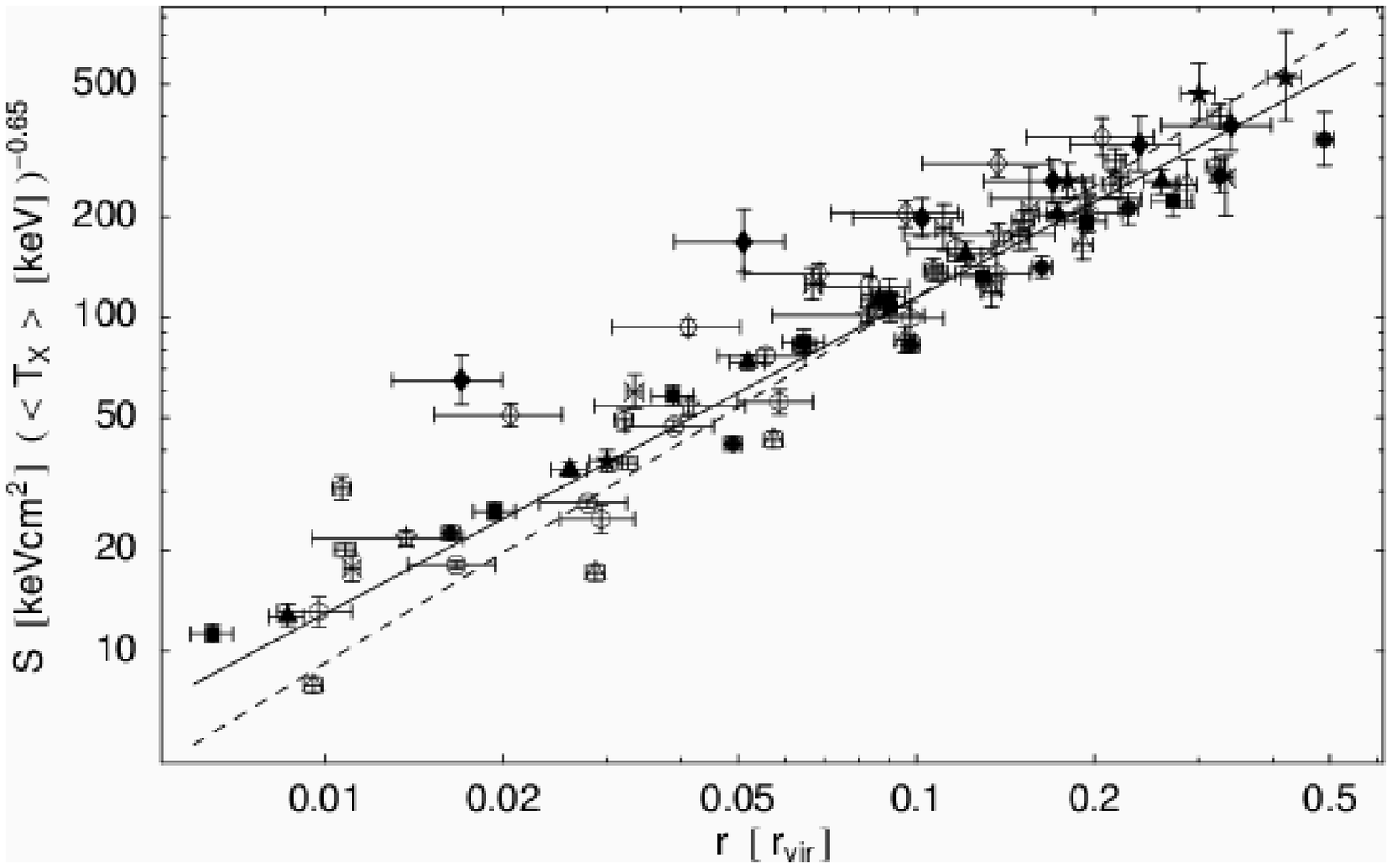}}
\caption{Same as Figure \ref{fig:entropynonscaled} but with the
scaling $S \propto h^{-4/3}(z) \, T^{0.65}$ used instead. The solid line indicates the best fit
power-law (best fit power index 0.95) and the dashed one the best fit
power-law with the power index value fixed to $1.1$.}
\label{fig:entropyscaled}
\end{figure*}

\begin{figure*}
\resizebox{\hsize}{!}{\includegraphics{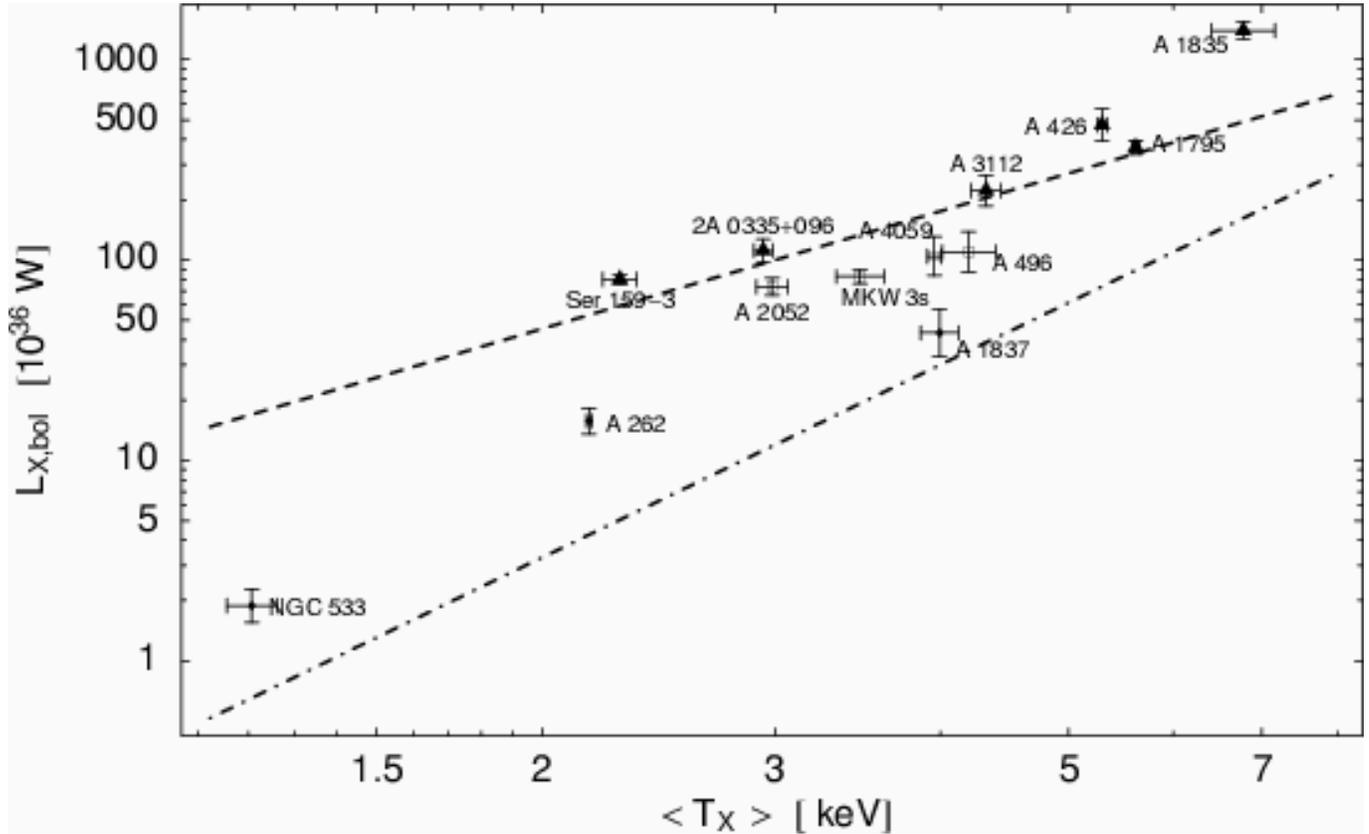}}
\caption{The cooling flow ``uncorrected'' $L$-$T$ relation in SCDM50. The plotted
quantities are the mean emission-weighted temperature
$<T_{\mathrm{X}}>$ and the bolometric X-ray luminosity within
$r_{500}$. {\sl Filled triangles} and {\sl open boxes} indicate
clusters with $\dot{M} > 300 \, \mathrm{M}_{\sun} \, \mathrm{yr}^{-1}$
and  $\dot{M} >
100 \, \mathrm{M}_{\sun} \, \mathrm{yr}^{-1}$, respectively. Mass
deposition rates for A~1837 are not available in the literature, while
$\dot{M}$ for A~262 and NGC~533 are consistent with zero. The dashed line shows
\cite{mccarthy04} predictions for an
entropy injection level of 200 keV cm$^2$, while the dot-dashed
line an entropy injection level of 500 keV cm$^2$.}   
\label{fig:Ltr500}
\end{figure*}

\begin{figure*}
\resizebox{\hsize}{!}{\includegraphics{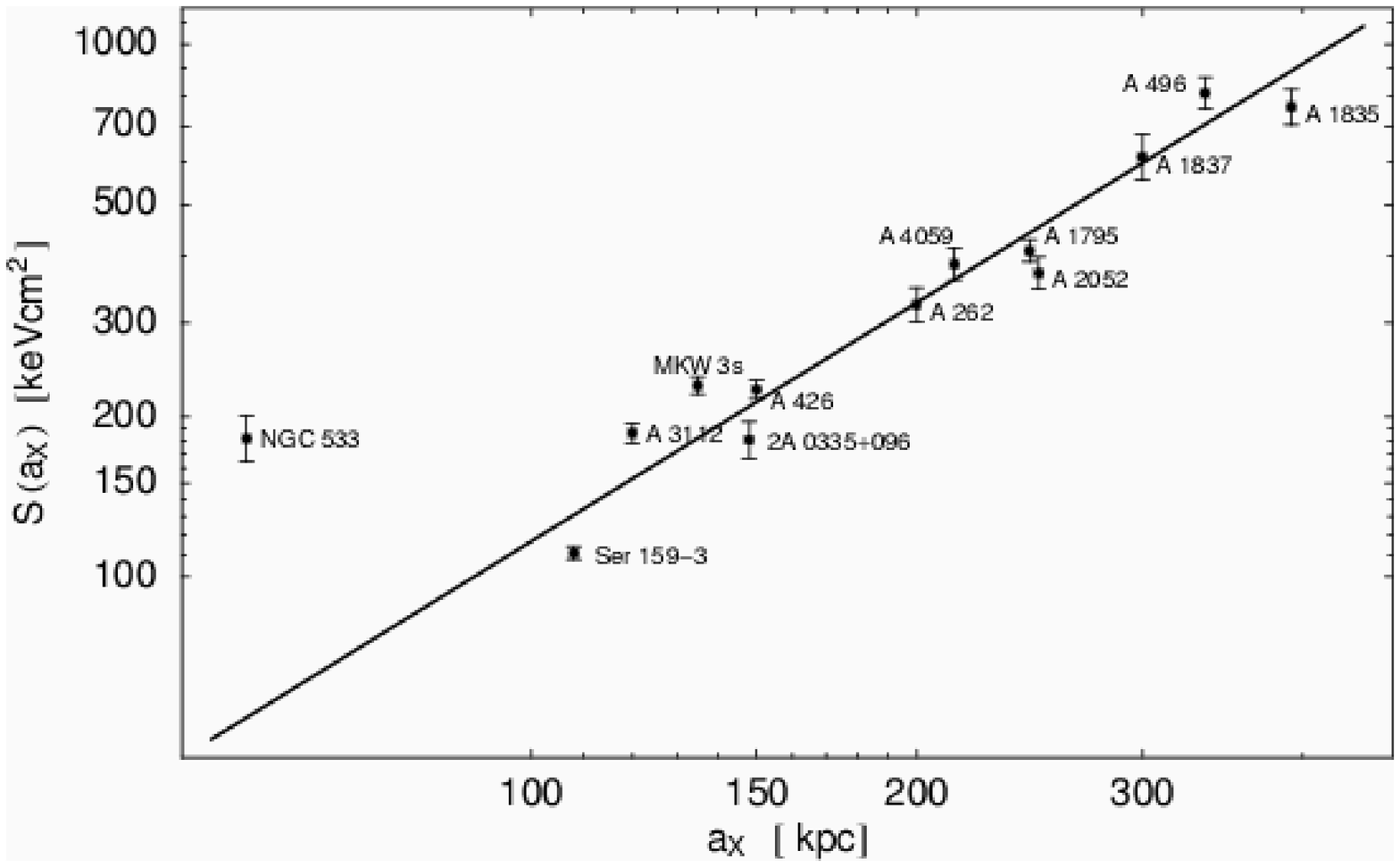}}
\caption{The core radius-central entropy relation in SCDM50. The solid line
indicates the best fit power law. NGC~533 is excluded from the fit.}    
\label{fig:Srcore}
\end{figure*}

The entropy is a useful quantity to probe the thermodynamic history
of the hot gas in groups and clusters. As it has become
customary, we define the entropy as $S=k
T/n_{\mathrm{e}}^{2/3}$, where $T$ and $n_{\mathrm{e}}$ are the
deprojected electron temperature and density, respectively. $S$, which
is by some authors referred as to 'entropy index', is related to the
thermodynamic entropy via a logarithm and an additive constant and
therefore conserved in any adiabatic process.\\* 
Particular interest has been
given to the shape of entropy profiles in groups, since preheating
models predict both large isentropic cores and a break in the entropy
temperature relation (e.g., \cite{tozzi01}). While there is evidence
that even at group scale no large isentropic cores are present
(\cite{pratt03}, \cite{ponman03}), the entropy profiles of
cooling flow clusters are expected to increase monotonically
moving outwards and to show no isentropic cores (e.g., \cite{mccarthy04}).\\*  In Fig.~\ref{fig:entropynonscaled} we plot entropy profiles for the
13 cooling flow clusters with the radial coordinate scaled by
$r_{\mathrm{vir}}$ in the SCDM50 cosmology. A visual inspection of the
individual profiles in the plot shows that, as
expected for systems with short cooling times in their central regions, no cores are present and the
entropy of the ICM increases monotonically moving outwards.\\* 
In the standard self-similar scenario, one predicts $S \propto
h^{-4/3}(z) \, T$, where $h^2(z)=\Omega_{\mathrm{m}} (1+z)^3
+\Omega_{\mathrm{\Lambda}}$ (e.g., \cite{pratt04}). On the other hand,
recent results by \cite{ponman03} suggest that entropy scales with
temperature as $S \propto T^{0.65}$, the
so-called ``entropy ramp'', instead of $S \propto T$. Both scalings $S \propto
h^{-4/3}(z) \, T$ and  $S \propto
h^{-4/3}(z) \, T^{0.65}$ considerably reduce the
dispersion of the profiles shown in
Fig.~\ref{fig:entropynonscaled}. Using $r_{\mathrm{vir}}$ to scale
radii, we quantify the dispersion of the scaled profiles using the standard deviation and the
mean of the scaled entropy values at a fixed scaled radius. As mean
cluster temperature $T$, the mean
emission-weighted temperature $<T_{\mathrm{X}}>$ (see
Table~\ref{tab:clusmasses}) is used. The scaled
entropy values are evaluated at fractions of $r_{\mathrm{vir}}$ for
which no extrapolation is needed and we compute the ratio between their
standard deviation $\sigma$ and mean $m$ to quantify the
relative dispersion of the scaled profiles. For both SCDM50 and
$\Lambda$CDM70 cosmologies we find that the dispersion is less if the
``entropy ramp'' scaling is used. For SCDM50 we find $\sigma / m =
0.37$ and $\sigma / m =
0.29$ at $0.1 r_{\mathrm{vir}}$ for the $S \propto T$ and $S \propto T^{0.65}$ temperature
scaling, respectively. The relative difference of the $\sigma / m $
ratios is the same in $\Lambda$CDM70 and the relative dispersion
$\sigma / m $ varies with radius only by a few
percent (e.g. $5 \%$ between $0.03$ and $0.2 r_{\mathrm{vir}}$ in
SCDM50). From scalings of the entropy profiles of
5 clusters (A~1983, A~2717, MKW9, A~1991 and A~1413), \cite{pratt04}
also find that
the scatter in the scaled entropy profiles is smaller if one uses \cite{ponman03}'s scaling rather than the standard self-similar
one. In addition to the dispersion analysis, we attempted to gain further
information on the scaling of entropy with temperature by directly
fitting the $S$-$T$ data (i.e. $S(0.1 \, r_{\mathrm{200}})$ versus
$<T_{\mathrm{X}}>$), but due to the small sample size and the fact
that the systems do not span a wide temperature range, we do not obtain any
statistically significant constrain on its slope. The profiles scaled using the relation $S \propto
h^{-4/3}(z) \, T^{0.65}$ are shown in Fig.~\ref{fig:entropyscaled}
for SCDM50. We fit the data shown in Fig.~\ref{fig:entropyscaled} with a line in log-log space (with
errors in both coordinates) and find, in excellent agreement with the
value found by \cite{pratt04}, a slope equal to $0.95 \pm
0.02$. Similarly, using {\sl Chandra} data, \cite{ettori02} find a slope equal
to $0.97$ for the entropy profile of A~1795. Therefore, our analysis gives additional evidence for a slope close to, but slightly shallower
than $1.1$, the value predicted by shock dominated spherical
collapse models (\cite{tozzi01}). This behavior is also a common
feature of numerical simulations (see \cite{pratt04} and references
therein). \\*
From the scaled entropy profiles we can obtain an estimate
of the normalization of the $S \propto T^{0.65}$ relation
found by \cite{ponman03}. For the entropy profiles with entropy values scaled as $S \propto
h^{-4/3}(z) \, (T/10 \, \mathrm{keV})^{0.65}$ and the radial
coordinate scaled by $r_{\mathrm{200}}$ we find a mean value at $0.1 \,
r_{\mathrm{200}}$ equal to 504 (with standard deviation 140). The
latter value is in excellent agreement with the normalization of
\cite{ponman03}'s $S(0.1 \, r_{\mathrm{200}})$-$T$ relation at $T=10
\, \mathrm{keV}$ (see their Figure 5). \\*  
Although the relation $S \propto T^{0.65}$ provides evidence for
similarity-breaking, the most robust evidence for the lack of self-similarity in the ICM
properties is provided by the luminosity-temperature ($L$-$T$) relation. The $L \propto
T^{\alpha}$ law, with $\alpha = 2$ expected for self-similar systems
is at variance with observations at both cluster and group scales. For
clusters, most authors (e.g. \cite{white97}, \cite{arnaud99}) have
found $\alpha \sim 3$, while for groups there is evidence for a much
steeper slope $\alpha \sim 5$ (e.g. \cite{helsdon00a,helsdon00b,xue00}). The latter result is however not
confirmed by the work of \cite{mulchaey98} and most important by the
recent analysis by \cite{osmond04}, as both works indicate
that the $L$-$T$ relation at group scales is consistent with and
extrapolation of the trend observed at cluster scale ( $\alpha \sim
3$).\\*
A great deal of semi-analytical and numerical studies have attempted to
give an explanation to the observed similarity-breaking. While it is
clear that models incorporating gravitationally-driven processes only
fail to match observations, there is evidence that models
incorporating a mixture of both cooling and heating processes are likely to
be successful (see \cite{ponman03} and references therein). Such
models should of course also explain the properties of cooling flow
clusters. Cooling flow clusters are known to enhance the scattering of the
observed $L$-$T$ relation, which is reduced once the
contribution from the cooling region is excised or cooling flow
clusters are excluded from the sample
(\cite{markevitch98,ettori02b}). On the other hand, the scatter in not
random and theoretical models incorporating gas cooling should be
compared to ``uncorrected'' data. This kind of comparison has never
been presented in semi-analytical or numerical works until recently in
\cite{mccarthy04}'s work. \cite{mccarthy04} have extended the work by
\cite{babul02} to include radiative cooling and compared their
predictions with observations without any ``cooling flow'' correction.\\*
In Fig. \ref{fig:Ltr500} we show the $L$-$T$ relation in SCDM50 for the 13 cooling
flow clusters we have analyzed. According to \cite{mccarthy04}'s predictions, clusters with large mass deposition rates $\dot{M}$ lie
on the high luminosity side of the scaling relation. We take mass deposition rates from \cite{peterson03} and, for the
objects not present in the latter work (A~3112 and A~426) from
\cite{peres}. Mass deposition rates for A~1837 are not available in
the literature. We divide the clusters into three groups: those with
$\dot{M} > 300 \, \mathrm{M}_{\sun} \, \mathrm{yr}^{-1}$ (``massive
cooling flow clusters''), with $\dot{M} >
100 \, \mathrm{M}_{\sun} \, \mathrm{yr}^{-1}$ (``moderate
cooling flow clusters'') and clusters with mass
deposition rates consistent with zero. Even though our sample is
small, a clear trend consistent with the findings of \cite{mccarthy04} can be
seen from a visual inspection of Fig. \ref{fig:Ltr500}, in which we
also show the predictions by \cite{mccarthy04} for an
entropy injection level of 200 keV cm$^2$ and 500 keV cm$^2$ (models
evolved for 10 Gyrs, see \cite{babul03}). ``Massive
cooling flow clusters'' are consistent with being systems that
experienced ``mild'' ($\sim$ 200 keV cm$^2$) levels of entropy
injection.\\*
We feel it should be pointed out that the entropy injection
levels in \cite{mccarthy04}'s model are quite high compared to most of the
preheating models (e.g., see \cite{voit2003} and references therein). However, a comparison between preheating
models is beyond the scope of our work, being our main purpose that of
providing ``uncorrected'' $L$-$T$ data for a sample of cooling flow clusters. 
In Fig. \ref{fig:Srcore} we show a second relationship predicted by
the work of \cite{mccarthy04}: the correlation between core radius and
central entropy. As expected, we find that the entropy at the core
radius $a_X$ increases with increasing core radius, with the exception
of the low mass system NGC~533. Using core radii taken from the
literature (see above) and excluding NGC~533 from the fit, we find:
$\mathrm{log}[S(a_X)]=1.48 \, \mathrm{log}[a_X] - 0.91$ (entropy in
keV cm$^2$ and core radius in kpc).

\section{Summary and Conclusions\label{sect:concl}}
We have investigated temperature and entropy
profiles of cooling flow clusters using the results from the spatially
resolved spectra taken with the EPIC cameras of XMM-Newton presented
in \cite{kaastra03}.\\*
Using deprojected plasma densities and temperatures we compute the total
gravitational mass profile and the virial radius of each clusters and find that:
\begin{itemize}
\item The inclusion of the
mass contribution from the gas in the estimate of the total mass of
the cluster is very important and gives more robust estimates.
\item As recently found by other authors, the estimated virial
radii substantially deviate from those estimated assuming the standard
size-temperature relation. 
\item Our total and gas mass estimates are in good agreement with
published ASCA and {\it BeppoSax} estimates. 
\end{itemize}
Our main findings concerning the temperature profiles are:
\begin{itemize}
\item When normalized and scaled by the virial radius, the temperature profiles are, except in the central cooling region, remarkably similar.
\item The profiles show a break radius $r_{\mathrm{br}}
\sim 0.12$ and $0.09 \, r_{\mathrm{vir}}$ in SCDM50
and $\Lambda$CDM70 cosmology, respectively.
\item Most important, the temperature profiles show a decline beyond the break
radius.
\item Our findings for the shape and slope of the profiles in this range is in
excellent agreement with previous studies by \cite{markevitch} and
\cite{dm02}. In SCDM50, a linear fit to the deprojected temperature
profiles with cluster-centric distance scaled by
the virial radius, gives a slope of $-0.97 \pm 0.14$ between $\sim
0.15 r_{\mathrm{vir}}$ and $0.50 r_{\mathrm{vir}}$.
\item For radii larger than $\sim 0.15
r_{\mathrm{vir}}$ (in SCDM50) most of the numerical simulations are able to
produce temperature profiles quite similar to the observed ones.
\item The temperature decline from $r_{\mathrm{br}}$ inwards is very
well parametrized using a profile with a core $r_{\mathrm{c}}$ (see
Eq. \ref{eqn:Tfit}). We
find that $r_c \propto <T_X>^{1.78\pm 0.10}$, with the mean cluster
temperature $<T_X>$ in keV and $r_c$ in kpc, and that $r_{\mathrm{c}}$ is not proportional to the cluster virial radius.
\end{itemize}
We have studied entropy profiles and find that:
\begin{itemize}
\item As expected for cooling flow clusters, i.e. massive systems with
strong cooling, no isentropic cores are present.
\item The entropy of the gas increases monotonically moving outwards
almost proportional to the radius.
\item The scatter in the scaled entropy profiles is smaller if the
empirical relation $S \propto T^{0.65}$ (\cite{ponman03}) is used
instead of the standard self-similar relation $S \propto T$. 
\item The profiles are close to, but slightly shallower ($\propto
r^{0.95}$ with entropies scaled using $S \propto T^{0.65}$) than those predicted by analytical models of shock dominated
spherical collapse ($\propto r^{1.1}$).
\item The normalization of the $S$-$T$ relation derived from the
scaling of the entropy profiles is in excellent agreement with the one
found by \cite{ponman03}.
\item We confirm the existence of a tight correlation between core
radius and central entropy.   
\end{itemize}


\begin{acknowledgements}
We wish to thank the referee for insightful comments which improved
the presentation of the results. This work is based on observations
obtained with XMM-Newton, an ESA science 
mission with instruments and contributions directly funded by 
ESA Member States and the USA (NASA). RP acknowledges partial support
by the Swiss National Science Foundation.
\end{acknowledgements}

\end{document}